***********************************************************

Article begins here

***********************************************************
\documentstyle[12pt]{article}
\newcommand{\beq}{\begin{equation}}
\newcommand{\eeq}{\end{equation}}
\newcommand{\beqa}{\begin{eqnarray}}
\newcommand{\eeqa}{\end{eqnarray}}
\newcommand{\labeq}[1]{\label{eq:#1}}
\newcommand{\refeq}[1]{\ref{eq:#1}}
\newcommand{\labref}[1]{\label{ref:#1}}
\newcommand{\rref}[1]{\ref{ref:#1}}

\newcommand{\bdm}{\begin{displaymath}}
\newcommand{\edm}{\end{displaymath}}
\newcommand{\bdma}{\begin{eqnarray*}}
\newcommand{\edma}{\end{eqnarray*}}

\begin{document}

\begin{flushright}
February 1992\\
DOE-288-CPP-26
\end{flushright}

\vspace*{24pt}

\begin{center}
{\large{\bf A Conformal Field Theory for the Quantum Hall Effect}}\\
\vspace{24pt}
Greg Nagao\\
Center for Particle Physics\\
The University of Texas\\
Austin, Texas~~78712
\end{center}
\vspace{64pt}

\centerline{\bf Abstract}
\vspace{12pt}

The QHE is studied in the context of a CFT.  An effective field of $N$
``spins" associated with the cyclotron motion of particles is taken as an
order parameter from which an effective Hamiltonian may be defined.  This
effective Hamiltonian describes the COM motion of the $N$ particles (with
coupling $\kappa_0$) together with a current-current interaction (of strength
$\kappa_1$).  Such a system gives rise to a CFT in the large $N$ limit when
$\kappa_0 = \kappa_1$.  The Laughlin wavefunction is derived from this CFT as
an $N'$-point correlation function of winding state vertex operators.

\newpage
\noindent
{\bf I.~~Introduction}
\vspace{12pt}

It has been suggested by Fubini [\rref{F2}] that the QHE might be
nicely understood through the use of CFT techniques.  In particular, Fubini
[\rref{F2}] and Stone [\rref{S1}] have outlined vertex operator
techniques which may be used to obtain the Laughlin wavefunction.  We show how
the vertex operators may be naturally defined in terms of an order parameter
arising from an effective Hamiltonian which includes a momentum-momentum
interaction between different particles [\rref{N2}]. This order parameter has
an interpretation of a field of ``bosonic spins" on scales much larger than
the magnetic length.  Under certain conditions (i.e. $\lim N \rightarrow
\infty$, $\kappa_0 = \kappa_1$, $\theta_i = \theta_j$ and radial ordering) the
$N$-particle system seems to exhibit a long-range order which may be
associated with the formation of winding states.  These winding states may be
characterized by its zero-modes representing the center of mass (COM) motion
of the system.  Such winding states transform as tensors under the group of
conformal transformations; the rank of the tensors being characterized by the
energy or angular momentum eigenvalues of the zero-modes.

The system is analagous to that of hydrodynamic turbulence in which the
assumption of scale invariance leads to a picture of ``eddies upon eddies".
These classical pictures are valid for special values of the
current-current couplings ($\kappa_0, \kappa_1$) in the theory.  They
correspond to points in field
space in which there exists a conformal symmetry.  As in string theories,
these fixed points correspond to different CFT's which describe ground states
of the system (i.e. classical solutions to the field equations).

In section II we define the collective coordinates which serve as our order
parameter for the theory.  From these, the winding state vertex operators of
Fubini's [\rref{F2}] are derived.  The $N'$ point correlation function
of the winding state vertex operators are then calculated in section III.  Here
it is shown that for special values of the coupling, this $N'$-point
correlation
function gives Laughlin's wavefunction in the large $N$-limit.  The anyonic
statistics of the ``quasiparticle" states created by these vertex operators is
then shown.  The quantization of the Hall conductance is then speculated to
arise from the topological properties of these winding states; associating the
winding states with solitons which wind around a compact configuration
space characterized by a two-dimensional modular parameter $\vec{L}$ (as for a
torus).  A discussion of the relation between the QHE and CFT is
then presented.

\vspace{24pt}
\noindent
{\bf II.~~Generalized Collective Coordinates for Winding States}
\vspace{12pt}

The collective coordinates of [\rref{N2}] may be generalized to include
arbitrary coefficients for the various modes of the fields

\beqa
\varphi_L (\zeta) &\equiv& \varphi_{0_L} + i \kappa_0 \alpha_0^* \ln \zeta +
\kappa_1 \sum_{n=1}^N \frac{1}{n} \left( \alpha_n \zeta^n + \alpha_n^*
\zeta^{-n} \right)\\
\varphi_R (\zeta^*) &\equiv& \varphi_{0_R} + i \kappa_0 \beta_0^* \ln \zeta^* +
\kappa_1 \sum_{n=1}^N \frac{1}{n} \left( \beta_n {\zeta^*}^n + \beta_n^*
{\zeta^*}^{-n} \right).
\eeqa

\noindent
The mode coefficients $\alpha_0$, $\alpha_n$, $\beta_0$ and $\beta_n$
may be viewed as annihilation-creation operators of a second quantized
theory governed by the $N$-particle Hamiltonian

\beqa
H &=& \frac{1}{2\mu} \left( \vec{\pi}_0^2 + \sum_{n=1}^N \vec{\pi}_n^2
\right)\\
\vec{\pi}_0 \equiv \vec{p}_0 + \frac{e}{c} \vec{A} (N \vec{x}_0) ~~~&;&~~~
\vec{\pi}_n \equiv \vec{p}_n + \frac{e}{c} \vec{A} (\vec{x}_n)\nonumber
\eeqa

\noindent
with the modes given by

\beqa
\alpha_0 \equiv \frac{\ell}{\sqrt{2N} \hbar} \left( \pi_0^1 + i \pi_0^2 \right)
 ~~~&;&~~~ \beta_0 \equiv \frac{\ell}{\sqrt{2N} \hbar} \left( \tilde{\pi}_0^2 +
i \tilde{\pi}_0^1\right)\nonumber\\
\alpha^*_0 \equiv \frac{\ell}{\sqrt{2N} \hbar} \left( \pi_0^1 - i \pi_0^2
\right) ~~~&;&~~~ \beta^*_0 \equiv \frac{\ell}{\sqrt{2N} \hbar} \left(
\tilde{\pi}_0^2 - i \tilde{\pi}_0^1\right)\nonumber\\
\alpha_n \equiv \frac{\ell}{\sqrt{2} \hbar} \left( \pi_n^1 + i \pi_n^2 \right)
 ~~~&;&~~~ \beta_n \equiv \frac{\ell}{\sqrt{2N} \hbar} \left( \tilde{\pi}_n^2 +
i \tilde{\pi}_n^1\right)\labeq{modes}\\
\alpha^*_n \equiv \frac{\ell}{\sqrt{2} \hbar} \left( \pi_n^1 - i \pi_n^2
\right)
 ~~~&;&~~~ \beta^*_n \equiv \frac{\ell}{\sqrt{2} \hbar} \left( \tilde{\pi}_n^2
-
i \tilde{\pi}_n^1\right)\nonumber\\
\vec{\tilde{\pi}}_0 \equiv \vec{p} - \frac{e}{c} \vec{A}(N\vec{x}_0)
 ~~~&;&~~~ \vec{\tilde{\pi}}_n \equiv \vec{p} - \frac{e}{c}
\vec{A}(\vec{x}_n)\nonumber
\eeqa

\noindent
where $\vec{x}_0 , \vec{p}_0$ are the COM position and momentum and
$\vec{\tilde{\pi}}$ is the bariocentric coordinate.  We have defined all but
the zero-mode commutation relation in previous papers
[\rref{N1},\rref{N2}].

Fubini [\rref{F2}] has shown that the Laughlin wavefunction may
be derived from vertex operators which appear to us to have the properties
of winding state vertex operators.  We shall show that this ideed is true.
Toward this end, we define the zero-modes $\varphi_{0_L}$ and
$\varphi_{0_R}$ in such a way that these commutation relations will be
proper to describe widning states, i.e.

\beq
\left[ \varphi_{0_L}, \alpha_0^\dagger \right] \sim -i ~~~~;~~~~ \left[
\varphi_{0_R}, \beta_0^\dagger \right] \sim -i.
\eeq

\noindent
This definition for the commutator will allow us to define winding state
vertex operators $V_\xi \sim e^{i \xi \varphi}$ (as in the usual formulation
of CFT) such that these states are eigenstates of the zero-mode operator
$\alpha_0^\dagger$.  We thus define the ``coordinate" zero mode as
\footnote{It is natural to introduce a factor of $\kappa_0$ for the COM zero
mode to annihilate $N$-particle state.}

\beqa
\varphi_{0_L} \equiv -i \alpha_0 \kappa_0 ~~~~~&;&~~~~~ \varphi_{0_R}
\equiv -i \beta_0 \kappa_0\labeq{zero modes}\\
\left[ \varphi_{0_L} , \alpha_0^\dagger \right] = -i \kappa_0 ~~~~~&;&~~~~~
\left[ \varphi_{0_R} , \beta_0^\dagger \right] = -i \kappa_0\nonumber
\eeqa

\noindent
We must now enlarge our Hilbert space to include both particle
andpsuedohole modes (such as in a closed-string theory).  A generalized
collective coordinate may then be defined as

\beq
\varphi ( \zeta, \zeta^*) \equiv \varphi_L (\zeta) + \varphi_R (\zeta^*).
\eeq

\noindent
The L(eft) and R(ight) subscripts denote the fact that while the L(eft) and
R(ight) fields are defined at the same positions ($x^1, x^2$) , they ``move" in
opposite directions under the influence of an external magnetic field due to
the fact that they are charge conjugates of each other.

Winding state vertex operators may then be defined as the normal ordered
product (denoted by $:~~:$) operators with different weightings for the
particles and holes

\beq
V_{\xi_L \xi_R} (\zeta, \zeta^*) \equiv : e^{i\left[ \xi_L \varphi_L (\zeta) +
\xi_R \varphi_R (\zeta^*) \right]} : .
\eeq

\noindent
If we define our Hilbert space in the usual way with respect to annihilation-
creation operators

\beqa
\alpha_0 |0> = 0 = \alpha_n |0> ~~~&;&~~~ <0| \alpha_0^\dagger = 0 = <0|
\alpha_n^\dagger\\
\beta_0 |0> = 0 = \beta_n |0> ~~~&;&~~~ <0| \beta_0^\dagger = 0 = <0|
\beta_n^\dagger,\nonumber
\eeqa

\noindent
we then see that the zero mode piece of these vertex operators form
eigenstates of these annihilation-creation operators (i.e.  they form coherent
states).  To see this, it is convenient to further separate our collective
coordinates into $\ell$(eft) and $r$(ight) - ordered pieces

\beqa
\varphi_L^\ell (\zeta) &\equiv & i \kappa_0 \alpha_0^\dagger \ln \zeta +
\kappa_1 \sum_{n=1}^N \frac{1}{n} \alpha_n^* \zeta^{-n}\nonumber\\
\varphi_L^r (\zeta) &\equiv & \varphi_{0_L} + \kappa_1 \sum_{n=1}^N
\frac{1}{n} \alpha_n \zeta^{-n}\\
\varphi_R^\ell (\zeta) &\equiv & i \kappa_0 \beta_0^\dagger \ln \zeta^* +
\kappa_1 \sum_{n=1}^N \frac{1}{n} \beta_n^* {\zeta^*}^{-n}\nonumber\\
\varphi_R^r (\zeta) &\equiv & \varphi_{0_R} + \kappa_1 \sum_{n=1}^N
\frac{1}{n} \beta_n {\zeta^*}^{-n}\nonumber
\eeqa

\noindent
so that the effect of our vertex operator on the ground state is

\beqa
<0| V_{\xi_L \xi_R} (\zeta , \zeta^*) &=& <0| e^{i\left[ \xi_L \varphi_L^\ell
(\zeta) + \xi_R \varphi_R^\ell (\zeta^*) \right]} e^{i\left[ \xi_L \varphi_L^r
(\zeta) + \xi_R \varphi_R^r (\zeta^*) \right]}\nonumber\\
  &=& <0| e^{i\left[ \xi_L \varphi_L^r (\zeta) + \xi_R \varphi_R^r (\zeta^*)
\right]}
\eeqa

\noindent
It is then easy to show that this is an eigenstate of $\alpha_0^\dagger$

\beqa
<0| V_{\xi_L \xi_R} (\zeta , \zeta^*) \alpha_0^\dagger &=& <0| e^{i\xi_L
\varphi_L^\ell (\zeta)} \left[ e^{i\xi_L \varphi_L^r(\zeta)}, \alpha_0^\dagger
\right]V_{\xi_R} (\zeta^*)\nonumber\\
 &=& <0| e^{i\xi_L \varphi_L^\ell (\zeta)} \xi_L e^{i\xi_L \varphi_L^r (\zeta)}
V_{\xi_R} (\zeta^*)
= <0| V_{\xi_L \xi_R} (\zeta, \zeta^*) \xi_L
\eeqa

\noindent
where

\beqa
\left[ e^{i\xi_L\varphi_L^r (\zeta)} , \alpha_0^\dagger \right] = e^{i\xi_L
\left( \varphi_L^r (\zeta) - \varphi_{0_L} \right)} \left[
e^{i\xi_L\varphi_{0_L}}, \alpha_0^\dagger \right] = \xi_L e^{i\xi_L
\varphi_L^r (\zeta)}.
\eeqa

\noindent
and

\beq
\left[ e^{i\xi_L\varphi_L^r (\zeta)}, \alpha_0^\dagger \right] = \kappa_0
\xi_L e^{i\xi_L\varphi_{0_L}}.
\eeq

\noindent
Likewise, it is an eigenstate of $\beta_0^\dagger$

\beq
<0| V_{\xi_L \xi_R} (\zeta,\zeta^*) \beta_0^\dagger = <0| V_{\xi_L \xi_R}
(\zeta,\zeta^*) \kappa_0 \xi_R.
\eeq

\vspace{24pt}
\noindent
{\bf III.~~Laughlin's Wavefunction from Winding State Vertex Operators}
\vspace{12pt}

Laughlin's $N'$ quasiparticle - quasihole wavefunction may now be derived
as a ``modular" ordered $N'$-point correlation function of winding state
vertex operators (apart from the conformal symmetry breaking exponential)

\beq
\Psi_{\xi_L \xi_R} \left( \zeta_1, \dots , \zeta_{N'}; \zeta_1^*, \dots ,
\zeta^*_{N'}\right) = e^{-\sum_{i=1}^{N'} |\zeta_i|^2} <0| M\left(
\prod_{i=1}^{N'} V_{\xi_L \xi_R} (\zeta_i, \zeta_i^*) \right) |0>
\eeq

\noindent
This is easily seen by contracting the $N'$-point correlation function:

\beqa
<0| M\left( \prod_{i=1}^{N'} V_{\xi_L \xi_R} (\zeta_i, \zeta_i^*) \right) |0>
&=& <0| M\left( \prod_{i<j=1}^{N'} \overbrace{V_{\xi_L \xi_R} (\zeta_i,
\zeta_i^*) V_{\xi_L \xi_R} (\zeta_j, \zeta_j^*) } \right) |0>\\
&=& <0|M\left( \prod_{i<j=1}^{N'} e^{-\xi_L^2\left [\varphi_L^r (\zeta_i),
\varphi_L^\ell (\zeta_j)\right]} e^{-\xi_R^2\left [\varphi_R^r (\zeta_i^*),
\varphi_R^\ell (\zeta_j^*)\right]} \right)|0>\nonumber
\eeqa

\noindent
where

\beq
\left[ \varphi_L^r(\zeta_i), \varphi_L^\ell (\zeta_j) \right] = \kappa_0^2 \ln
\zeta_j + \kappa_1^2 \sum_{n=1}^N \frac{1}{n} \left( \frac{\zeta_i}{\zeta_j}
\right)^n
\eeq

\noindent
and we have changed to ``ordinary" oscillators strictly for computational
ease.\footnote{We could have performed all calculations in this ordinary
oscillator basis as well.} The relation between these two different oscillators
is given by

\beqa
\alpha_n \equiv \sqrt{n} a_n ~~~;~~~ & \alpha_n^* \equiv \sqrt{n} a_n^* &
 ~~~;~~~ \alpha_{-n} \equiv \sqrt{n} a_{-n}\nonumber\\
\beta_n \equiv \sqrt{n} a_n ~~~;~~~ & \beta_n^* \equiv \sqrt{n} a_n^* &
 ~~~;~~~ \beta_{-n} \equiv \sqrt{n} a_{-n}\\
\left[a_n, a_m^\dagger \right] = & \delta_{nm} & = \left[ b_n, b_m^\dagger
\right].\nonumber
\eeqa

\noindent
Now, in the large $N$ limit, the series converges to a logarithm if we take
$\theta_i = \theta_j$ and $r_i < r_j$ for $\zeta \equiv r e^{i\theta}$.  Also
taking $\kappa_0 = \kappa_1$, we then have

\beqa
\lim_{N\rightarrow \infty} \lim_{\theta_i \rightarrow \theta_j}
\lim_{\kappa_1 \rightarrow \kappa_0} \left[ \varphi_L^r (\zeta_i),
\varphi_L^\ell (\zeta_j) \right]_{r_i < r_j} = \kappa_0^2 \ln(\zeta_j -
\zeta_i)\labeq{III.10}\nonumber\\
\lim_{N\rightarrow \infty} \lim_{\theta_i \rightarrow \theta_j}
\lim_{\kappa_1\rightarrow \kappa_0} \left[ \varphi_L^r (\zeta_i^*),
\varphi_L^\ell (\zeta_j^*) \right]_{r_i < r_j} = \kappa_0^2 \ln(\zeta_j^* -
\zeta_i^*).\labeq{III.11}
\eeqa

\noindent
To see this, note that

\beqa
\left[\varphi_L^r(\zeta_i), \varphi_L^\ell (\zeta_j)\right] &=& \kappa_0^2
\ln \zeta_j + \kappa_1^2 \sum_{n=1}^N \frac{1}{n} \left(
\frac{\zeta_i}{\zeta_j} \right)^n \nonumber\\
 &=& \kappa_0^2 \ln \zeta_j + \kappa_1^2 \sum_{n=1}^N \frac{1}{n}
\left(\frac{r_i}{r_j}\right)^n e^{in(\theta_i-\theta_j)}
\eeqa

\noindent
such that, if $\theta_i = \theta_j$, $r_i < r_j$ and $N \rightarrow \infty$,
the
sum converges to a logarithm

\beq
\sum_{n=1}^\infty \frac{1}{n} \left(\frac{r_i}{r_j} \right)^n = \ln \left( 1 -
\frac{r_i}{r_j} \right) = \ln \left( 1 - \frac{\zeta_i}{\zeta_j} \right),
\eeq

\noindent
giving

\beq
\left[\varphi_L^r(\zeta_i), \varphi_L^\ell (\zeta_j)\right] = \kappa_0^2 \ln
\zeta_j + \kappa_1^2 \ln \left( 1 - \frac{\zeta_i}{\zeta_j} \right)
\eeq

\noindent
which reduces to (\refeq{III.10}) if $\kappa_0 = \kappa_1$ (we have the
similar limit for the anti-analytic piece).

Our $N'$-point correlation function then becomes

\beqa
\lim_{N\rightarrow \infty} <0| M\left( \prod_{i=1}^{N'} V_{\xi_L
\xi_R} (\zeta_i, \zeta_i^*) \right)_{\theta_i = \theta_j}^{\kappa_1 =\kappa_0}
|0> &=&\prod_{i<j=1}^{N'} e^{-\xi_L^2 \kappa_0^2 \ln (\zeta_j - \zeta_i)} e^{-
\xi_R^2 \kappa_0^2 \ln(\zeta_j^* - \zeta_i^*)}\nonumber\\
 &=& \prod_{i<j=1}^{N'} \left( \zeta_j - \zeta_i \right)^{-\kappa_0^2 \xi_L^2}
\left( \zeta_j^* - \zeta_i^* \right)^{-\kappa_0^2 \xi_R^2}.
\eeqa

\noindent
Here we see that the conformal weights of these vertex operators is given by

\beq
\left(h,\tilde{h}\right)  = \left( \kappa_0^2 \xi_L^2 , \kappa_0^2 \xi_R^2
\right)
\eeq

\noindent
where $\kappa_0$ plays the role of the slope parameter $\alpha'$ in a
closed string theory.

Thus, Laughlin's wavefunction may be obtained as a coherent state in a
triple limit  of an $N$-particle Hamiltonian with momentum-momentum
coupling between different particles.  In this triple limit (i.e. the large
$N$-
limit, with all vertex operators aligned along the same angle ($\theta_i =
\theta_j~~~\forall~~i,j=1,\dots, N'$) and for special values of the couplings
($\kappa_0 = \kappa_1$), such a state corresponds to a modularly ordered
$N'$-point correlation function of winding state vertex operators.  It then
transforms as a tensor of rank ($\kappa_0^2\xi_L^2, \kappa_0^2 \xi_R^2$)
under the group of two-dimensional conformal transformations whose field-
theoretical Hamiltonian may be taken as

\beq
H = \hbar \omega \left( \kappa_0^2 \alpha_0^\dagger \alpha_0 +
\kappa_1^2 \sum_{n=1}^N \alpha_n^\dagger \alpha_n \right).
\eeq

\noindent
In the case that $\kappa_0 \neq \kappa_1$, the ``equal angle" commutator
between fields may be written (in the large $N$-limit) as

\beqa
\left[\varphi_L^r(\zeta_i), \varphi_L^\ell (\zeta_j) \right]_{\theta_i =
\theta_j}^{N\rightarrow \infty} &= & \kappa_0^2 \ln \zeta_j + \kappa_1^2
\frac{1}{n} \left( \frac{\zeta_i}{\zeta_j}\right)^n\nonumber\\
 &=& \left(\kappa_0^2 - \kappa_1^2 \right) \ln \zeta_j + \kappa_1^2
\ln(\zeta_j - \zeta_i)
\eeqa

\noindent
The $N'$-point correlation function then becomes

\beq
<0| M\left( \prod_{i=1}^{N'} V_{\xi_L \xi_R} (\zeta_i, \zeta_i^*) \right) |0> =
\prod_{i<j=1}^N (\zeta_j - \zeta_i)^{-\kappa_1^2 \xi_L^2} (\zeta_j^*-
\zeta_i^*)^{\kappa_1^2 \xi_R^2} |\zeta_j|^{2(\kappa_0^2 - \kappa_1^2)}
\eeq

\noindent
which we may identify as the wavefunction for a quasiparticle and
quasipsuedohole
fluctuation (or excitation) centered at $\zeta = 0$
[\rref{F2},\rref{L3},\rref{PG}].

\vspace{24pt}
\noindent
{\bf IV.~~Statistics of the Winding States}
\vspace{12pt}

To determine the statistics of the winding states, we consider the rotation of
one ``quasiparticle" around another in the Laughlin wavefunction.  In the
modularly ordered (i.e. radially ordered) correlation function, we define
$\zeta \equiv re^{i\theta}$ such that $r_1 < r_2$ and ${\zeta '}_2 \equiv
\zeta_2 e^{2\pi i}$.  The unrotated two- quasiparticle wavefunction may
then be written as

\beqa
\psi_{\xi_L\xi_R} (\zeta_1, \zeta_2 ;\zeta_1^*, \zeta_2^*) &\sim& <0| M \left(
V_{\xi_L \xi_R} ({\zeta }_1 , {\zeta }_1^*) V_{\xi_L \xi_R} ({\zeta }_2 ,
{\zeta }_2^*) \right) |0>\nonumber\\
&\sim& e^{\xi_L^2 \left[\varphi_L^r(\zeta_1), \varphi_L^\ell(\zeta_2)\right]}
e^{\xi_R^2 \left[\varphi_R^r(\zeta_1^*) , \varphi_R^\ell(\zeta_2^*)\right]}
\nonumber\\
&\sim& e^{\kappa_0 \xi_L^2 \ln (\zeta_2 - \zeta_1)} e^{\kappa_0 \xi_R^2 \ln
(\zeta_2^* - \zeta_1^*)}
\eeqa

\noindent
where the commutators are ``equal angle" commutators taken in the large
$N$-limit.  On the other hand, we have for the rotated vertex operator
$V_{\xi_L \xi_R} ({\zeta '}_2 , {\zeta '}_2^*)$ the ``equal angle" commutators

\beqa
\xi_L^2 \left[ \varphi_L^r(\zeta_1), \varphi_L^\ell({\zeta'}_2)
\right]_{\theta_1 = \theta_2} &=&
\xi_L^2 \left[ \kappa_0^2 \ln {\zeta '}_2 + \kappa_1^2 \sum_{n=1}^N
\frac{1}{n} \left(\frac{\zeta_1}{{\zeta '}_2} \right)^n\right]\nonumber\\
 &=& \xi_L^2 \left[ \kappa_0^2 (\ln \zeta_2 + 2\pi i ) + \kappa_1^2
\sum_{n=1}^N \frac{1}{n} \left(\frac{\zeta_1}{\zeta_2}\right)^n  e^{-2 \pi in}
\right]\nonumber\\
 &=& \xi_L^2 \left[ \varphi_L^r(\zeta_1), \varphi_L^\ell({\zeta}_2)
\right]_{\theta_1 = \theta_2}
+ 2\pi i \kappa_0^2 \xi_L^2\\
\xi_R^2 \left[ \varphi_R^r(\zeta^*_1), \varphi_R^\ell({\zeta '}^*_2)
\right]_{\theta_1 = \theta_2} &=&
\xi_R^2 \left[ \varphi_R^r(\zeta^*_1), \varphi_R^\ell({\zeta }^*_2)
\right]_{\theta_1 = \theta_2} -
2\pi i \kappa_0^2 \xi_R^2\nonumber
\eeqa

\noindent
giving rise to the rotated wavefunction

\beq
\psi_{\xi_L \xi_R} (\zeta_1, {\zeta '}_2 ; \zeta_1^*, {\zeta '}_2^* ) =
\psi_{\xi_L \xi_R} (\zeta_1, {\zeta }_2 ; \zeta_1^*, {\zeta }_2^* ) e^{2\pi i
\kappa_0^2 \left(\xi_L^2 - \xi_R^2 \right)}.
\eeq

\noindent
This leads to anyonic statistics for the quasiparticles associated with the
coherent state of the winding vertex operator.  The
wavefunction is not required to be single-valued; however particles
described by vertex operators with phases which change by integer
amounts, i.e. satisfying the condition

\beq
\kappa_0^2 \left(\xi_L^2 - \xi_R^2 \right) \in Z
\eeq

\noindent
describe particles of ``ordinary" statistics.  Such states of course give rise
to single-valued wavefunctions.  If one requires locality and closure of the
operator product algebra, one can
further show that these vertex operators must have conformal weights such
that $(\xi_L, \xi_R) \in \Gamma$ where $\Gamma$ is a lattice of Lorentzian
signature.

\vspace{24pt}
\noindent
{\bf V.~~An Interpretation for the Winding States}
\vspace{12pt}

As we have just seen in the previous section, the statistics of the
quasiparticle created by the winding state vertex operator is given by the
difference between the conformal weights of its analytic and anti-analytic
factors.  This follows from the fact that the rotation operator is given by
[\rref{N2}]

\beq
L_3 = \hbar (L_0 - \tilde{L}_0 )
\eeq

\noindent
The phase change can be attributed to a shift in the zero-mode contributing
to the COM angular momentum.

To see this explicitly, we calculate the effect of a rotation of our order
parameter $\varphi(\zeta_2, \zeta_2^*)$ around the vertex operator
$V_{\xi_L \xi_R} (\zeta_1 ,\zeta_1^*)$ in the radially ordered product with
$r_1 < r_2$.  For ${\zeta '}_2 = \zeta_2 e^{2\pi i}$, we then have

\beqa
<0| V_{\xi_L} (\zeta_1) \varphi_L ({\zeta '}_2 ) &=& <0|V_{\xi_L} (\zeta_1)
\left[ \varphi_{0_L} + i \kappa_0 \alpha_0^\dagger \ln {\zeta '}_2 +
{\sum_{n=-N}^N}^{\textstyle \prime} \alpha_n {\zeta '}_2^n \right]\\
 &=& <0|V_{\xi_L} (\zeta_1) \left[ \varphi_{0_L} - 2 \pi \kappa_0 \xi_L + i
\kappa_0 \alpha_0^\dagger \ln {\zeta }_2 + \kappa_1
{\sum_{n=-N}^N}^{\textstyle \prime} \alpha_n {\zeta }_2^n e^{2\pi i n}
\right]\nonumber\\
 &=& <0| V_{\xi_L} (\zeta_1) \left[ \varphi_L ({\zeta }_2 ) - 2\pi \kappa_0
\xi_L \right]\nonumber\\
<0| V_{\xi_R} (\zeta_1^*) \varphi_R ({\zeta '}_2^* ) &=& <0| V_{\xi_R}
(\zeta_1^*) \left[ \varphi_R ({\zeta '}_2^* ) + 2\pi \kappa_0 \xi_R
\right]
\eeqa

\noindent
which gives the overall effect on the vertex operator

\beq
<0| V_{\xi_L \xi_R} (\zeta_1, \zeta_1^*) \varphi_R ({\zeta '}_2,
{\zeta '}_2^* ) = <0| V_{\xi_L \xi_R} (\zeta_1, \zeta_1^*) \left[ \varphi_R
({\zeta }_2, {\zeta }_2^*) + 2 \pi \kappa_0 (\xi_R - \xi_L)\right]
\eeq

\noindent
where the prime over the sum ($\sum^\prime$) indicates that omission of the
zero mode in the
sum. Thus, the effect is to shift the zero modes

\beqa
\varphi_{0_L} &\longrightarrow& \varphi_{0_L} - 2\pi \kappa_0 \xi_L
\labeq{left shift}\\
\varphi_{0_R} &\longrightarrow& \varphi_{0_R} - 2\pi \kappa_0 \xi_R
\labeq{right shift}.
\eeqa

\noindent
 From equations (\refeq{modes},\refeq{zero modes})

\beqa
\varphi_{0_L} &=& \frac{\kappa_0 \ell}{\sqrt{2N}\hbar} \left(-i \pi_0^1 +
\pi_0^2 \right)\nonumber\\
\varphi_{0_R} &=& \frac{\kappa_0 \ell}{\sqrt{2N}\hbar} \left(-i
\tilde{\pi}_0^2 + \tilde{\pi}_0^1 \right).\nonumber\\
\eeqa

\noindent
The shift in the zero mode (\refeq{left shift},\refeq{right shift}) may be
absorbed into $\pi_0^2$ and $\tilde{\pi}_0^1$, giving

\beqa
\pi_0^2 &\longrightarrow& \pi_0^2 - k_L = p_0^2 + \frac{e}{c}A^2(N\vec{x}_0) -
k_L\labeq{eqn:V.1}\\
\tilde{\pi}_0^1 &\longrightarrow& \pi_0^1 + k_R = p_0^1 - \frac{e}{c}
A^1(N\vec{x}_0) + k_R\nonumber
\eeqa

\noindent
where $k_L \equiv 2\pi \xi_L \sqrt{2N} \frac{\hbar}{\ell}$ and $k_R \equiv
2\pi \xi_R \sqrt{2N} \frac{\hbar}{\ell}$.  Then in arbitrary gauge, the winding
state leads to a shift in the COM momentum

\beqa
p_0^2 &\longrightarrow& p_0^2 - k_L\\
p_0^1 &\longrightarrow& p_0^1 + k_R\nonumber
\eeqa

\noindent
This leads to a shift in the COM angular momentum

\beq
x_0^1p_0^2 - x_0^2p_0^1 \longrightarrow x_0^1p_0^2 - x_0^2p_0^1 - \Delta
L_3^{(0)}
\eeq

\noindent
where $\Delta L_3^{(0)} \equiv x_0^1 k_L + x_0^2 k_R$.  The winding state
vertex operator $V_{\xi_L \xi_R} (\zeta,\zeta^*)$ has the effect of creating a
quasiparticle fluctuation whose COM energy is shifted from that of the
original ground state.

An alternative description can also be made in coordinate space.  Rather than
absorbing the shift into the momentum, we may absorb it into the vector
potential which can be written in arbitrary gauge as

\beq
A^i(\vec{x}) = \frac{B}{2} \left(\epsilon^{ij} x^j + \frac{\partial
\Lambda}{\partial x^i} \right).
\eeq

\noindent
 From (\refeq{eqn:V.1}) we then have

\beqa
A^2(N\vec{x}_0) &\longrightarrow& A^2(N\vec{x}_0) - \frac{c}{e} k_L\\
A^1(N\vec{x}_0) &\longrightarrow& A^1(N\vec{x}_0) - \frac{c}{e} k_R\nonumber
\eeqa

\noindent
which, in symmetric gauge ($\Lambda = 0$) gives

\beqa
x_0^1 &\longrightarrow& x_0^1 + L_0^1\\
x_0^2 &\longrightarrow& x_0^2 + L_0^2\nonumber
\eeqa

\noindent
where $L_0^1 \equiv 4\pi \xi_L \ell \sqrt{\frac{2}{N}}$ and $L_0^2 \equiv
-4\pi \xi_R \ell \sqrt{\frac{2}{N}}$.  Our overall zero-
mode then chages by the amount

\beqa
\varphi_0 \equiv \varphi_{0_L} + \varphi_{0_R} &\longrightarrow&
\varphi_0 + 2\pi \kappa_0 (\xi_R - \xi_L)\nonumber\\
\Longrightarrow \vec{x}_0 &\longrightarrow& \vec{x}_0 + \vec{L}_0.
\eeqa

If we require the wavefunction to be single-valued, we must have

\beq
\kappa_0 (\xi_R - \xi_L) = M ~~~~~;~~~~~ {\rm for}~M\in Z
\eeq

\noindent
or, in terms of the real coordinate zero-modes (i.e. COM position)

\beq
\vec{x}_0 \sim \vec{x}_0 + \vec{L}_0 \labeq{equivalence}
\eeq

\noindent
suggesting that the winding mode be associated with the cyclotron motion of
a quasiparticle.  The equivalence relation defined by (\refeq{equivalence})
suggests that $\vec{x}_0 \in \bar{\cal M}$ where $\bar{\cal M}$ is some
compact manifold characterized by the independent ``winding (modular)
parameters"
composing $\vec{L}_0$.  For a two-quasiparticle wavefunction there will
only be two such independent parameters; however for an $N'$-quasiparticle
wavefunction

\beq
\psi_{\xi_{L_1} \dots \xi_{L_{N'}}; \xi_{R_1} \dots \xi_{R_{N'}}} \left(
\zeta_1, \dots ,\zeta_{N'}; \zeta_1^*, \dots,\zeta^*_{N'} \right) \sim <0|
M\left(
\prod_{i=1}^{N'} V_{\xi_{L_i} \xi_{R_i}} (\zeta_i,\zeta_i^* ) \right) |0>
\eeq

\noindent
the number of parameters will obviously be related to the number of
independent phases associated with the rotations of vertex operators around
each other.  The dimension of $\vec{L}_0$ thus increases with $N'$.  For $N'
= 1$, $\bar{\cal M}$ will be diffeomorphic to $S^1 \times S^1 \sim T^2$.  The
field $\varphi (\zeta , \zeta^*) = \varphi_L (\zeta) + \varphi_R(\zeta^*)$ can
thus be recognized as fields coordinatizing a group manifold $\bar{\cal M}$
where

\beqa
g_L(\xi_L) \equiv e^{i\xi_L \varphi_L} \in G_L ~~~~&;&~~~~ g_R(\xi_R) \equiv
e^{i\xi_R \varphi_R} \in G_R\nonumber\\
g(\xi_L,\xi_R) \equiv g_L(\xi_L) g_R(\xi_R) ~~~~&;&~~~~ g(\xi_L,\xi_R) \in G
\sim G_L \times G_R
\eeqa

\noindent
may be recognized as our vertex operators which are elements of the group
$G$.  This has obvious generalizations to include other interactions (such as
spin-spin interactions) in the context of the nonlinear sigma model (NLSM)
and Kac-Moody algebras via the Wess-Zumino-Witten theory.\footnote{We
expect the conductivity tensor to be associated with the symmetric and
antisymmetric couplings of the NLSM.  The idea is to relate a metric and
torsion tensor on the group manifold to the longitudinal and Hall components
of the conductivity tensor; the torsion being related to the Hall
conductivity.}

\vspace{24pt}
\noindent
{\bf VI.~~The QHE as a CFT}
\vspace{12pt}

In the previous sections we have shown how the Laughlin wavefunction
could be derived by defining an order parameter $\varphi (\zeta , \zeta^*)$
 from which a winding state vertex operator could be defined.  Under the
special conditions of

\begin{itemize}
\item the large $N$-limit

\item modular (or radial) ordering

\item equal-angle commutators

\item and special values for the couplings $\kappa_1 = \kappa_0$
\end{itemize}

\noindent
we found that one can obtain the Laughlin $N'$-quasiparticle wavefunction
as an $N'$-point correlation function of winding state vertex operators (apart
 from an exponential factor which can be attributed to a boundary term).
This leads to the picture of the quantum Hall effect as a manifestation of
some critical phenomena expressible in the language of CFT.

To better understand the relations of the QHE with CFT, we may view our
order parameter $\varphi(\zeta, \zeta^*)$ as that of a field of spins (created
by $\alpha^\dagger$) and ``anti-spins" (created by $\beta^\dagger$).  This
picture is valid in the low-energy limit - that is at energies $E<< \hbar
\omega$ in which case the cyclotron motion of the electrons is may be taken
as time-independent (the oscillators $\alpha^\dagger$ create energy
eigenstates and are thus associated with stationary states).  At higher
energies, the spins degrees of the electrons must then be taken into account.
In a very strong magnetic field, however, the electron spins are all aligned
with the external magnetic field (i.e. all electrons are
polarized)\footnote{The
spin-flip energy is much larger than that of the gap energy of the Landau
levels.  We expect this energy to be of order $e^2$ (like the fine structure
constant for spin interactions in an atom) as opposed to the order $e$
interaction of the cyclotron motion ($\omega = \frac{eB}{\mu c}$).}  We
expect different effective field theories to be applicable at different
energies.
 From the perspective of the NLSM, the couplings\footnote{In this case we
may take our couplings to be associated with the components of the
conductivity tensor $\sigma_{ij} (\varphi (\zeta, \zeta^*))$.} of the effective
field theory become renormalized at lower and lower energies according to
the renormalization group (RG) equations.  Fixed points of these RG equations
correspond to critical couplings associated with those of a CFT.  These points
are viewed as local minima of the effective potential.  As a local minimum of
an effective potential, the field configurations correspond to those which
minimize the action, i.e. they correspond to field equations for a classical
field theory.  In connection with integrable systems, these classical field
theories may be associated with ``particle-like" configurations due to a
coherent superposition of the many particle system creating a quasiparticle
state (i.e. bound state configuration).  Here we have seen an example in
which the winding state vertex operators create such coherent states which
are eigenstates of the COM complex momentum $\alpha_0$.  The
integrability of the system is due to the large (infinite) number of conserved
quantities which commute with the Hamiltonian at the critical point.  The
long range order associated with the conformal invariance may thus be
related to the $N'$-quasiparticle coherent state formed by the $N'$-point
correlation function of winding state vertex operators (in the triple limit
previously describe) which gives rise to the Laughlin wavefunction
describing this $N'$-quasiparticle state.  It may also be viewed as that long
range interaction responsible for the current-current coupling which has led
us to this formulation of the QHE in terms of a CFT [\rref{N2}].

A CFT may be characterized by a stress-energy tensor built out of the order
parameters of the theory.  In particular, the Sugawara construction utilizes
the conserved currents such that the stress-energy tensor is quadratic in
these currents.  Such is the case with our construction here.  We may define
the stress-energy tensor as

\beqa
T(\zeta, \zeta^*) &\equiv& T_L (\zeta) + T_R (\zeta^*)\nonumber\\
T_L(\zeta) \equiv  \frac{\partial \varphi_L(\zeta)}{\partial \zeta}
\frac{\partial \varphi_L(\zeta)}{\partial \zeta} ~~~&;&~~~ T_R(\zeta^*) \equiv
 \frac{\partial \varphi_R(\zeta^*)}{\partial \zeta^*} \frac{\partial
\varphi_R(\zeta^*)}{\partial \zeta^*}.
\eeqa

\noindent
Here we may recognize the conserved current as that associated with our
conjugate momentum [\rref{N2}]

\beq
j_L(\zeta) \equiv \frac{\partial \varphi_L(\zeta)}{\partial \zeta} = \pi
(\zeta) ~~~~;~~~~
j_R(\zeta^*) \equiv \frac{\partial \varphi_R(\zeta^*)}{\partial \zeta^*} =
\tilde{\pi}(\zeta^*).
\eeq

The generators of the conformal transformations under which the theory is
invariant may be obtained from the moments of the stress-energy tensor.
The resulting Virasoro algebra consists of two commuting pieces associated
with the analytic and anti-analytic terms (i.e. L(eft) and R(ight) pieces
associated with the particles and psuedoholes, respectivly).  Expanding our
stress-energy tensors, we have

\beqa
T_L(\zeta) &=& \frac{1}{\zeta^2} \left[ -\kappa_0^2 \alpha_0^\dagger
\alpha_0 + i \kappa_0 \kappa_1 \alpha_0^\dagger \sum_{n=1}^N
\left(\alpha_0 \zeta^n - \alpha_n^* \zeta^{-n} \right) \right.\nonumber\\
 & & \left. + \kappa_1^2 \sum_{n,m=1}^N \left( \alpha_n \alpha_m
\zeta^{n+m} + \alpha_n^* \alpha_m^* \zeta^{n+m} - \alpha_n^* \alpha_m
\zeta^{m-n} \right) \right]\nonumber\\
 &\equiv& - \sum_{n=-N}^N \frac{L_n}{\zeta^{n+2}}
\eeqa

\noindent
giving the generator of dilations (in the large $N$ limit with $\kappa_0 =
\kappa_1$)

\beq
L_0 = \kappa_0^2 \alpha_0^\dagger \alpha_0 + \kappa_1^2 \sum_{n=1}^N
\left(\alpha_n \alpha_n^* + \alpha_n^* \alpha_n \right).
\eeq

\noindent
Likewise for our ``psuedoholes"

\beq
\tilde{L}_0 \equiv \kappa_0^2 \beta_0^\dagger \beta_0 + \kappa_1^2
\sum_{n=1}^N \left(\beta_n \beta_n^* + \beta_n^* \beta_n \right).
\eeq

\noindent
Normal ordering rather than symmeterizing the dilation operator gives the
usual dilation operator.  The effective Hamiltonian is then given by

\beqa
H (\kappa_0, \kappa_1) \equiv H_L(\kappa_0, \kappa_1) &+& H_R
(\kappa_0, \kappa_1)  = \hbar \omega (L_0 + \tilde{L}_0) \nonumber\\
H_L(\kappa_0, \kappa_1) \equiv \hbar \omega L_0(\kappa_0, \kappa_1)
 ~~~&;&~~~ H_R(\kappa_0, \kappa_1) \equiv \hbar \omega
\tilde{L}_0(\kappa_0, \kappa_1).
\eeqa

\noindent
To emphasize the fact that this Hamiltonian corresponds to that of a CFT only
for special values of the couplings (and in the large $N$ limit), we have
denoted the coupling dependence explicitly.

\newpage
\noindent
{\bf VII.~~Conclusion}
\vspace{12pt}

Our results suggest that the Laughlin wavefunction describes a state of the
system corresponding to a critical point in a phase transition such that it is
describable by a CFT.  In our example model (which excludes electron spins),
the current-current coupling seems to be the source of the long-range order of
the system.  Cast in the more general context of a 2+1 dimensional NLSM,
we expect the effective field theory to be related to a chiral symmetry
breaking which gives rise to the QHE.

Many connections between the QHE, CFT and strings still remain to be
uncovered.  Among the more interesting connections of interest is the
connection between the Landau levels and Regge trajectories.  This requires a
better understanding of such things as the physical state conditions and its
relation to the measurements of the conductivity tensor.

\vspace{48pt}
{\bf Acknowledgements}
\vspace{12pt}

It is a pleasure to thank Prof. Q. Niu for many useful discussions.  This work
was supported in part by a Dept. of Energy grant \# DOE-288-CPP-26.

\newpage
\begin{center}
{\large {\bf References}}
\end{center}

\begin{enumerate}



\item Fubini S., {\em Vertex Operators and the Quantum Hall Effect},
CERN-TH. 5922/90.\\
Fubini S. and L\"{u}tken C.A., {\em Vertex Operators in the Fractional
Quantum Hall Effect}, CERN-TH. 5960/90.\labref{F2}

\item Stone M., {\em Vertex Operators in the Quantum Hall Effect}, IL-
TH-90-\# 32.\labref{S1}




\item Nagao G., {\em Toward a CFT for the Quantum Hall Effect},
DOE-287-CPP-25.\labref{N2}

\item Nagao G., {\em Psuedoholes, Gauge Constraints and Critical Phenomena in
the Quantum Hall Effect}, unpublished.\labref{N1}

\item Laughlin R.B., Phys. Lett. {\bf 50} (1983) 1395.\labref{L3}

\item {\em The Quantum Hall Effect}, ed. by Prange and Girvin.\labref{PG}

\end{enumerate}

\end{document}